# Analysis of ChatGPT on Source Code

Ahmed R. Sadik[1], Antonello Ceravola[1], Frank Joublin[1], Jibesh Patra[1]

[1] Honda Research Institute Europe

**Abstract**. This paper explores the use of Large Language Models (LLMs) and in particular ChatGPT in programming, source code analysis, and code generation. LLMs and ChatGPT are built using machine learning and artificial intelligence techniques, and they offer several benefits to developers and programmers. While these models can save time and provide highly accurate results, they are not yet advanced enough to replace human programmers entirely. The paper investigates the potential applications of LLMs and ChatGPT in various areas, such as code creation, code documentation, bug detection, refactoring, and more. The paper also suggests that the usage of LLMs and ChatGPT is expected to increase in the future as they offer unparalleled benefits to the programming community.

# Contents

# 1 Introduction

The role of the human programmer has been vital throughout the development of machine computing. Initially, programmers had to manually code instructions for computers to follow. This was a time-consuming and error-prone process. As computers became more powerful, programming languages were developed that were high-level and made it easier to write code. This resulted in the creation of the first code generators. Over the years additional tools and subsystems have been created to assist human programmers, leading to design patterns, several development methodologies, high-level debuggers, intelligent editors, and code generators. However, the human programmer remained responsible for creating and maintaining the software. In recent years, there has been a growing interest in the democratization of software development. Two main streams are the no-code paradigms (based on model-driven-development methods) and AI generated code. In this paper we investigate the usage of applications of AI in assisting various software engineering tasks.

Machine learning and artificial intelligence algorithms have been developed in a way that they can analyze large datasets of source code and generate source code that resembles human written code. While such approaches have the potential to make programming more accessible and efficient, they are not yet advanced enough to replace the human programmer entirely. One example of a widely used AI system, for software development, made available since 2021 is Copilot, an AI-powered pair programming tool integrated in GitHub [Bird et Al. 2022][Zaremba et Al. 2021]. Copilot quickly got attention from a large number of developers, and it has been integrated in several software development tools. In this paper we investigate the recent research on LLMs and their capabilities in supporting the software development process in different ways [Gozalo-Brizuela et Al. 2023].

We show here several tests done with ChatGPT [OpenAI 2022] and other related research results available in existing literature, with the purpose of inspiring and giving ideas for further work with AI in the scope of software development in the scientific community [van Dis 2023]. Although most of the examples here are done with ChatGPT, they are not limited to this LLM. Several other language models are available, sharing some characteristics and/or providing some original ones that could be exploited for achieving specific software development tasks or as support for other domains.

The use of LLMs for programming, source code analysis, and code generation is a revolutionary development in the field of computer science. Progress in this area has shown numerous innovations, unthinkable a few years ago by a community of reseachers looking at the limitation of deep neural network. For instance one example of that can be found in the opinions of AI practitioners such as Chollet, the creator of the popular deep learning library Keras. As a limitation of deep learning, Chollet writes, *"…many more applications are completely out of reach for current deep learning techniques—even given vast amounts of human annotated data. Say, for instance, that you could assemble a dataset of hundreds of thousands—even millions—of English language descriptions of the features of a software product, as written by a product manager, as well as the corresponding source code developed by a team of engineers to meet these requirements. Even with this data, you could not train a deep learning model to simply read a product description and generate the appropriate codebase. That's just one example among many…"* [Chollet 2017]. The current developments and the results shown in this white paper, demonstrates that such claims were too bold and that new deep learning architecture like transformer open new territories for deep learning also in discrete space like language or coding models.

# 2 Model Training

## 2.1 Background

On March 15, 2022, OpenAI unveiled the latest iterations of their groundbreaking GPT-3 and Codex models in their API. Dubbed "text-davinci-003" and "code-davinci-002", these models revealed impressive new capabilities for editing and insertion, surpassing the previous versions in terms of power and versatility. The models had been trained on an extensive dataset up until June 2021, ensuring that they were at the cutting edge of language and coding AI.

OpenAI continued to push the boundaries of AI language models with the introduction of a new series on November 30, 2022: "GPT-3.5". This series included the aforementioned text and code models, which were further fine-tuned for even greater performance. In addition, OpenAI introduced ChatGPT, a model that was specifically fine-tuned from a GPT-3.5 variant. With this new addition to their product line, OpenAI cemented their position as leaders in the field of natural language processing and artificial intelligence.

## 2.2 CODEX

The Codex models are descendants of GPT-3 models that can understand and generate code. Their training data contains both natural language and billions of lines of public code from GitHub. Codex is most capable in Python and proficient in over a dozen languages including JavaScript, Go, Perl, PHP, Ruby, Swift, TypeScript, SQL, and even Shell. Codex training, fine tuning, and accuracy was described in detail in the article "Evaluating Large Language Models Trained on Code" [Chen et Al. 2022].

At the time of writing this paper (March 2023), [Chen et Al. 2022] article is the only available information that can trace the methods ChatGPT has been developed with. GPT-NEO, GPT-J, and GPT-3 are among the most advanced autoregressive language models available today, utilizing deep learning algorithms to generate text that closely resembles human writing. These models work by taking an initial text prompt and then continuing the text in a manner that is coherent and contextually relevant. However, the capabilities of these models are not limited to generating natural language text. Researchers have found that GPT-3 has the ability to generate simple programs from text strings, even though it was not specifically trained to do so. This observation has led to the hypothesis that a specialized variant of GPT-3 called Codex would excel at a variety of coding tasks, given its enhanced capabilities for generating code. As the field of artificial intelligence continues to evolve, it is clear that models like Codex will play an increasingly important role in shaping the future of natural language processing and machine learning.

## 2.3 Method

Codex is a powerful AI language model that was specifically designed for coding tasks. It is pre-trained on an enormous amount of raw data, including 54 million GitHub repositories and 179 GB of Python files, each under 1 MB in size. To ensure the quality of the training data, OpenAI filtered out auto-generated code based on criteria such as average and maximum line length and a small number of alphabetic characters. After filtering, the clean data set consisted of 159 GB of Python files. Codex is trained on a variable number of parameters, ranging from 300 million to 12 billion, depending on the specific use case. The vast amount of raw data and careful filtering process have enabled Codex to develop an advanced understanding of the structure and syntax of code in various programming languages, making it an incredibly powerful tool for a wide range of coding tasks.

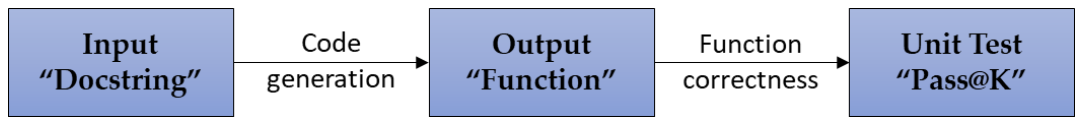

*Figure 1: the evaluation method of Codex is most probably the same method used to evalute ChatGPT*

Codex has taken a different approach to evaluate the quality of its output, unlike other language models that use match-based matrices like BLEU score to measure natural language quality. Function correctness is used to measure the quality of Codex's output as it is a more effective measure in the context of code. Unlike natural language, code has a different structure with a limited precise vocabulary. BLEU score has problems capturing the code's semantics as it's not well-suited for the diverse and ambiguous vocabulary of code. Codex, on the other hand, uses function correctness, which is used by human developers to judge the quality of code. The idea behind this approach is that software requirements should be converted into test cases before implementation begins, and success is determined by a program that passes these tests. While not all organizations employ test-driven development, integration of new code typically depends on creating and passing unit tests, making function correctness a valuable metric for evaluating the quality of Codex's output.

## 2.4 Training

Codex was trained using a similar approach to the GPT model. The model was trained with a linear warmup of 175 steps and cosine learning rate decay, using Adam optimizer with a total of 100 billion tokens.. This training process allowed Codex to learn the complex patterns and structures of code by adjusting its parameters to optimize performance on a wide range of coding tasks. The approach used in training Codex reflects the significant computational resources and careful consideration required to build such a powerful AI language model, capable of producing high-quality code for a variety of applications.

## 2.5 Evaluation

To evaluate the functional correctness of Codex, a set of 164 programming problems was used, called the HumanEval dataset. Each problem included a function signature, docstring, body, and multiple unit tests, with an average of 7.7 tests per problem. The tasks were carefully hand-written to assess language comprehension, reasoning, algorithms, and basic mathematics. It was essential for these tasks to be manually created because Codex was trained on a large portion of GitHub, which already contained solutions to problems from various sources. For example, there were more than ten public repositories containing solutions to Codeforces problems, which were part of the recently proposed APPS dataset. Therefore, it was crucial to use a unique and carefully constructed dataset to evaluate the functional correctness of Codex accurately. The HumanEval dataset served as an effective tool to evaluate the model's ability to perform a diverse range of coding tasks, providing insights into its strengths and weaknesses.

Table 1: Codex, GPT-Neo, & TabNine evaluations for HumanEval.

|   | PASS@$k$ | | |
|---|---|---|---|
|   | $k=1$ | $k=10$ | $k=100$ |
| GPT-NEO 125M | 0.75% | 1.88% | 2.97% |
| GPT-NEO 1.3B | 4.79% | 7.47% | 16.30% |
| GPT-NEO 2.7B | 6.41% | 11.27% | 21.37% |
| GPT-J 6B | 11.62% | 15.74% | 27.74% |
| TABNINE | 2.58% | 4.35% | 7.59% |
| CODEX-12M | 2.00% | 3.62% | 8.58% |
| CODEX-25M | 3.21% | 7.1% | 12.89% |
| CODEX-42M | 5.06% | 8.8% | 15.55% |
| CODEX-85M | 8.22% | 12.81% | 22.4% |
| CODEX-300M | 13.17% | 20.37% | 36.27% |
| CODEX-679M | 16.22% | 25.7% | 40.95% |
| CODEX-2.5B | 21.36% | 35.42% | 59.5% |
| CODEX-12B | 28.81% | 46.81% | 72.31% |

To evaluate the performance of Codex on programming problems, multiple solution samples were generated from the models and tested to see if they passed the unit tests. With just a single sample, the 12B parameter Codex solves 28.8% of these problems, and the 300M parameter Codex solves 13.2% of these problems. In comparison, the 6B parameter GPT-J only achieves 11.4% on the same dataset, and all GPT models achieve near 0%. To further improve Codex's ability to synthesize functions from docstrings, the model was fine-tuned on standalone, correctly implemented functions, resulting in Codex-S, which can solve 37.7% of problems with a single sample. In real-world programming tasks, multiple approaches and bug fixes are often required, which is approximated by generating many samples from the models and selecting one that passes all unit tests. Codex-S was able to generate at least one correct function within 100 samples for 77.5% of the problems, suggesting that accurate code samples can be selected via heuristic ranking. In fact, the sample with the highest mean log-probability passes unit tests for 44.5% of the problems.

## 3 Applications

Research on the applications of the LLMs for Software Development has shown a myriad of possibility. This include, code generation, documentation generation, bug detection, code smell detection, refactoring of existing code, translating one programming language to another, test case generation, debugging, code clone detection, program synthesis, code summarization to name a few. In this paper, we focus on the following applications of LLMs on source code. We use ChatGPT for the purpose of our investigation and provide supporting examples for each such application.

- Code smells detection an refactoring
- Code generation
- Step-by-Step execution of source code
- Bug fixing

- Code comprehension improving
- Bad coding practices detection and refactoring
- Code analysis and summarization

For a more comprehensive introduction on LLMs in the scope of this paper, refer to [Joublin et Al. 2023].

## 3.1 Refactoring

Code smells are specific patterns or characteristics in code that indicate the presence of a deeper problem. These problems may include design flaws, inefficiencies, or issues with maintainability or readability. Code smells do not necessarily indicate a bug or error in the code, but they can make it harder to understand, modify, and maintain. Common examples of code smells include duplicate code, long methods, unused variables, and overly complex class hierarchies. Identifying and addressing code smells is an important part of the software development process, as it can improve code quality, reduce bugs, and make it easier to maintain and extend the codebase over time. Refactoring code smells are indicators that the code may benefit from refactoring to improve its quality.

### 3.1.1 Bloaters

These smells refer to a type of code smells that refer to code that has grown too large and complex, making it difficult to maintain, understand, and modify. There are five types of Bloaters code smells:

#### 3.1.1.1 Long parameter list

More than three or four parameters in a method call

```
PROMPT:
this function suffers from long parameters list, can you fix!

def calculate_pizza_price(size, dough_type, toppings, delivery):
    base_price = get_base_price(size)
    dough_price = get_dough_price(dough_type)
    toppings_price = get_toppings_price(toppings)
    delivery_price = get_delivery_price(delivery)
    return base_price + dough_price + toppings_price + delivery_price

pizza_price = calculate_pizza_price("large", "thick", ["mushrooms", "pepperoni"], True)
print(pizza_price)

ChatGPT:
Yes, we can fix the issue of a long parameter list in the calculate_pizza_price function by using a dictionary to pass in the arguments instead:

def calculate_pizza_price(pizza):
    base_price = get_base_price(pizza['size'])
    dough_price = get_dough_price(pizza['dough_type'])
    toppings_price = get_toppings_price(pizza['toppings'])
    delivery_price = get_delivery_price(pizza['delivery'])
    return base_price + dough_price + toppings_price + delivery_price

pizza = {'size': 'large', 'dough_type': 'thick', 'toppings': ['mushrooms', 'pepperoni'], 'delivery': True}
pizza_price = calculate_pizza_price(pizza)
print(pizza_price)
```

Long parameters code smell is fuzzy concept for ChatGPT, as "long" is relative. Therefore, the programmer needs to explicity ask ChatGPT to refactor the long parameters codesmell. Otherwise, ChatGPT will not be able to discover it. The solution that ChatGPT has used to fix the long parameters codesmell is very practicle as it guarantee the scalablity of the parameter list, and the encapsulation of the Pizza class.

### 3.1.1.2 Long method

A long method is hard to understand. The key to make it easy to understand is not the method length but naming. You should not read the actual implementation to understand what a method does.

```
PROMPT:
this functions is long, can you fix it!

def bake_and_deliver_pizza(size, toppings, dough_type, name, address):
    pizza = make_pizza_with_ingredients(size, toppings, dough_type)
    if not pizza:
        return "Error: Pizza could not be made"
    else:
        bake_pizza(pizza)
        if not check_doneness(pizza):
            return "Error: Pizza not done"
        else:
            package_pizza(pizza)
            delivery_status = deliver_pizza(name, address, pizza)
            if delivery_status:
                return f"Pizza delivered to {name} at {address}."
            else:
                return "Error: Pizza could not be delivered."
ChatGPT:
Yes, the bake_and_deliver_pizza function violates the Single Responsibility Principle because it is responsible for multiple tasks:

Making the pizza
Baking the pizza
Checking if the pizza is done
Packaging the pizza
Delivering the pizza
Instead, these responsibilities should be separated into different functions or classes with each having a single responsibility.

Here's an example of how the bake_and_deliver_pizza function can be refactored to follow the SRP:

def make_pizza(size, toppings, dough_type):
    pizza = Pizza(size, toppings, dough_type)
    return pizza

def bake_pizza(pizza):
    baker = PizzaBaker()
    baker.bake(pizza)

def check_doneness(pizza):
    checker = PizzaChecker()
    return checker.check_doneness(pizza)

def package_pizza(pizza):
    packager = PizzaPackager()
    packager.package(pizza)

def deliver_pizza(name, address, pizza):
    delivery = PizzaDelivery(pizza, name, address)
    return delivery.deliver()

def bake_and_deliver_pizza(size, toppings, dough_type, name, address):
    pizza = make_pizza(size, toppings, dough_type)
    if not pizza:
        return "Error: Pizza could not be made"
    bake_pizza(pizza)
    if not check_doneness(pizza):
        return "Error: Pizza not done"
    package_pizza(pizza)
    delivery_status = deliver_pizza(name, address, pizza)
    if delivery_status:
        return f"Pizza delivered to {name} at {address}."
    else:
        return "Error: Pizza could not be delivered."
```

ChatGPT can detect the code smell within the given function as the question was clearly asked about the specific code smell. ChatGPT has used the Single Responsibility Principle (SRP) concept to devide the long method into smaller ones.

### 3.1.1.3 Large class

Large classes are hard to read, understand, and maintain.

```
PROMPT:
this class is long, can you fix it!

  def __init__(self, size, toppings, dough_type, name, address):
        self.size = size
        self.toppings = toppings
        self.dough_type = dough_type
        self.name = name
        self.address = address

    def assemble_pizza(self):
        dough = make_dough(self.dough_type)
        sauce = add_sauce()
        toppings = add_toppings(self.toppings)
        return dough, sauce, toppings

    def bake_pizza(self):
        # bake the pizza
        return "Pizza is baked"

    def deliver_pizza(self):
        # simulate delivery
        return f"Pizza delivered to {self.name} at {self.address}"

ChatGPT:
Yes, the Pizza class violates the Single Responsibility Principle (SRP) because it has multiple responsibilities:
Assembling the pizza
Baking the pizza
Delivering the pizza
Instead, these responsibilities should be separated into different classes with each class having a single responsibility:
A PizzaAssembler class responsible for assembling the pizza.
A PizzaBaker class responsible for baking the pizza.
A PizzaDelivery class responsible for delivering the pizza.
By separating these responsibilities into different classes, we can achieve better maintainability, testability, and flexibility in our code.

PROMPT:
can you apply the SRP by dividing the pizze class into two classes. First class to make pizza, second class to deliver it

ChatGPT:
class PizzaMaker:
    def __init__(self, size, toppings, dough_type):
        self.size = size
        self.toppings = toppings
        self.dough_type = dough_type

    def assemble(self):
        assembler = PizzaAssembler(self.dough_type, self.toppings)
        return assembler.assemble()
    def bake(self):
        baker = PizzaBaker()
        return baker.bake()
class PizzaDelivery:
    def __init__(self, pizza, name, address):
        self.pizza = pizza
        self.name = name
        self.address = address

    def deliver(self):
        delivery = PizzaDeliveryService(self.pizza, self.name, self.address)
        return delivery.deliver()
```

ChatGPT can detect the code smell within the given class, as the question was clearly asked about this specific code smell. In this case it is important to ask ChatGPT to perform a specific solution by dividing the class into two classes, as many solution for this problem can be used. Therefore, the user should be able to understand the problem and know the solutions to manipulate the right answer.

### 3.1.1.4  Data Clumps

Two or more data fields are frequently used together in a class or structure and should be grouped together.

```
PROMPT:
this code suffers from data clumps, can you fix it?

class PizzaOrder:

def calculate_pizza_price(size, dough_type, toppings, delivery_location, delivery_time):
    base_price = get_base_price(size)
    dough_price = get_dough_price(dough_type)
    toppings_price = get_toppings_price(toppings)
    delivery_price = get_delivery_price(delivery_location, delivery_time)
    return base_price + dough_price + toppings_price + delivery_price

size = "large"
dough_type = "thick"
toppings = ["mushrooms", "pepperoni"]
delivery_location = "New York"
delivery_time = "evening"
pizza_price = calculate_pizza_price(size, dough_type, toppings, delivery_location, delivery_time)
print(pizza_price)

ChatGPT:
Yes, in this code, the calculate_pizza_price function takes 5 parameters: size, dough_type, toppings, delivery_location, and delivery_time.
These parameters are all related to the pizza order and they are often passed together in the same function calls. This is an example of
data clumps. One possible solution is to group related data into a single data structure such as a dictionary or a class. Here is an
example of how to refactor the code using a dictionary. in this example, we create a dictionary pizza that contains all the relevant data
for calculating the pizza price. The calculate_pizza_price function takes in the pizza dictionary as an argument and uses the dictionary
keys to retrieve the relevant data. This way, we avoid passing multiple arguments to the function, which can be cumbersome and prone to
errors.

By using a dictionary, we have grouped related data together and made the code more organized and easier to read.

class PizzaOrder:

def calculate_pizza_price(pizza):
    base_price = get_base_price(pizza['size'])
    dough_price = get_dough_price(pizza['dough_type'])
    toppings_price = get_toppings_price(pizza['toppings'])
    delivery_price = get_delivery_price(pizza['delivery_location'], pizza['delivery_time'])
    return base_price + dough_price + toppings_price + delivery_price

pizza = {
    'size': "large",
    'dough_type': "thick",
    'toppings': ["mushrooms", "pepperoni"],
    'delivery_location': "New York",
    'delivery_time': "evening"
}

pizza_price = calculate_pizza_price(pizza)
print(pizza_price)
```

ChatGPT can detect the code smell within the given code, as the question was clearly asked about this specific code smell. ChatGPT provides one good possible solution by encapsulating the Pizza parameters into a dictionary structure.

### 3.1.1.5 *Primitive Obsession*

Primitive refers to primitive data types. Obsession is to use these data types instead of objects/classes to represent domain ideas.

```
PROMPT:
this code suffers from primitive obsession, can you fix it?

def calculate_pizza_price(size, dough_type, toppings):
    base_price = get_base_price(size)
    dough_price = get_dough_price(dough_type)
    toppings_price = get_toppings_price(toppings)
    return base_price + dough_price + toppings_price

size = "large"
dough_type = "thick"
toppings = "mushrooms, pepperoni"
pizza_price = calculate_pizza_price(size, dough_type, toppings)
print(pizza_price)

ChatGPT:
Yes, in this code, the calculate_pizza_price function takes 3 parameters as string type, that are all related to the pizza order. The
problem with this approach is that the function takes string as input and it can be misused, for example the size can be written in small
letter, the dough type can be written in plural form, and toppings are written as string, and it can be written in different format.

from enum import Enum
from typing import List
class PizzaSize(Enum):
    SMALL = "small"
    MEDIUM = "medium"
    LARGE = "large"
class PizzaDoughType(Enum):
    THIN = "thin"
    THICK = "thick"
class PizzaToppings(Enum):
    MUSHROOM = "mushrooms"
    PEPPERONI = "pepperoni"
    CHEESE = "cheese"
class Pizza:
    def __init__(self, size: PizzaSize, dough_type: PizzaDoughType, toppings: List[PizzaToppings]):
        self.size = size
        self.dough_type = dough_type
        self.toppings = toppings

the above code addresses the primitive obsession issue by defining enums for pizza size, dough type, and toppings, and using them as
parameters for the Pizza class constructor. This makes the code more readable and maintainable by abstracting away the details of pizza
size, dough type, and toppings and providing a clear and defined interface for creating a Pizza object.
```

Primitive obsession code smell is a fuzzy concept for ChatGPT, as "obsession" is relative. Therefore, the programmer needs to explicity ask ChatGPT to refactor the primitive obsession codesmell. Otherwise, ChatGPT will not be able to discover it. ChatGPT provides one good viable solution defining enums.

### 3.1.2 Couplers

These smells refer to a set of bad coding practices that involve classes that are highly dependent on each other. Coupling refers to the degree of interdependence between two classes, and it is desirable to keep coupling low, as high coupling can make the system harder to understand, maintain, and modify.

Couplers code smells are a type of design problem that indicates a high level of coupling between classes. To avoid couplers code smells, it is important to keep classes loosely coupled and to use design patterns that promote low coupling, such as the Dependency Injection pattern. These smells include:

#### *3.1.2.1 Inappropriate Intimacy*

Classes have too much knowledge of each other's internal workings, and changes in one class require changes in another class. For example a method in one class frequently accesses the methods of another class.

```
PROMPT:
this code suffers from inappropriate intimacy, can you fix it?

class Pizza:
    def __init__(self, size, dough_type):
        self.size = size
        self.dough_type = dough_type

    def calculate_price(self):
        base_price = get_base_price(self.size)
        dough_price = get_dough_price(self.dough_type)
        return base_price + dough_price

class Order:
    def __init__(self, pizza):
        self.pizza = pizza

    def calculate_price(self):
        return self.pizza.calculate_price() + self.pizza.dough_type == "thick"

pizza = Pizza("large", "thick")
order = Order(pizza)
order_price = order.calculate_price()
print(order_price)

ChatGPT:
Yes, the code suffers from inappropriate intimacy because the Order class is directly accessing the Pizza class's dough_type attribute,
which breaks the principle of encapsulation. To fix this, we can add a method to Pizza to return the dough_type value and remove the
direct access to the attribute in Order

Now, the Order class is using the get_dough_type method of the Pizza class to retrieve the dough_type value instead of directly accessing
the attribute. This way, we maintain encapsulation and avoid the inappropriate intimacy code smell.

class Pizza:
    def __init__(self, size, dough_type):
        self.size = size
        self.dough_type = dough_type

    def calculate_price(self):
        base_price = get_base_price(self.size)
        dough_price = get_dough_price(self.dough_type)
        return base_price + dough_price

    def get_dough_type(self):
        return self.dough_type

class Order:
    def __init__(self, pizza):
        self.pizza = pizza

    def calculate_price(self):
        return self.pizza.calculate_price() + (self.pizza.get_dough_type() == "thick")

pizza = Pizza("large", "thick")
order = Order(pizza)
order_price = order.calculate_price()
print(order_price)
```

ChatGPT can detect the code smell within the given code, as the question was clearly asked about this specific code smell. ChatGPT provides one good possible solution by adding a get_dough_type method that decouble the pizza and the order classes.

### *3.1.2.2 Feature Envy*

A method in one class frequently accesses the data of another class.

---

**PROMPT:**
this code suffers from **Feature Envy**, can you fix it?

```python
class Pizza:
    def __init__(self, size, dough_type):
        self.size = size
        self.dough_type = dough_type

    def calculate_price(self):
        base_price = get_base_price(self.size)
        dough_price = get_dough_price(self.dough_type)
        return base_price + dough_price

class Order:
    def __init__(self, pizza):
        self.pizza = pizza

    def calculate_price(self):
        return self.pizza.calculate_price() + self.pizza.dough_type == "thick"

pizza = Pizza("large", "thick")
order = Order(pizza)
order_price = order.calculate_price()
print(order_price)
```

**ChatGPT:**
Yes, in this code, the **Order class has a method calculate_price that calculates the total price of the order, which involves calling the calculate_price method of the Pizza class and then adding the delivery price**.

the most straightforward solution to fix the inappropriate intimacy issue in this code **is to add a is_thick method to the Pizza class that returns a boolean indicating whether the pizza's dough type is thick or not**, and then modify the Order class to call this method instead of directly accessing the dough_type member variable. Here's an example:

```python
class Pizza:
    def __init__(self, size, dough_type):
        self.size = size
        self.dough_type = dough_type

    def calculate_price(self):
        base_price = get_base_price(self.size)
        dough_price = get_dough_price(self.dough_type)
        return base_price + dough_price

    def is_thick(self):
        return self.dough_type == "thick"

class Order:
    def __init__(self, pizza):
        self.pizza = pizza

    def calculate_price(self):
        if self.pizza.is_thick():
            return self.pizza.calculate_price() + 2.0
        else:
            return self.pizza.calculate_price()

pizza = Pizza("large", "thick")
order = Order(pizza)
order_price = order.calculate_price()
print(order_price)
```

---

ChatGPT can detect the code smell within the given code, as the question was clearly asked about this specific code smell. ChatGPT provides one good possible solution by adding a is_thick method that decouble the pizza and the order classes.

### 3.1.2.3 *Message Chains*

A series of method calls are chained together in a single statement.

```
PROMPT:
this code suffers from Message Chains, can you fix it?

class Pizza:
    def __init__(self, size, dough_type):
        self.size = size
        self.dough_type = dough_type

    def calculate_price(self):
        base_price = get_base_price(self.size)
        dough_price = get_dough_price(self.dough_type)
        return base_price + dough_price

class Order:
    def __init__(self, pizza):
        self.pizza = pizza

    def calculate_price(self):
        return self.pizza.calculate_price() + self.pizza.dough_type == "thick"

pizza = Pizza("large", "thick")
order = Order(pizza)
order_price = order.calculate_price()
print(order_price)

ChatGPT:
Yes, there is a message chain issue in this code. **A message chain occurs when an object sends a message to another object, which then
sends a message to yet another object**, and so on, resulting in a long chain of method calls.

In this case, the Order class is calling the calculate_price method of the Pizza class twice in the calculate_price method of the Order
class, which creates a message chain. This could make the code harder to read and understand, and could also lead to performance issues
if the chain involves many objects or method calls.

To fix this issue, you could simply **call the calculate_price method of the Pizza class once and store the result in a variable**, and then
use that variable to calculate the final order price. Here's an example:

class Pizza:
    def __init__(self, size, dough_type):
        self.size = size
        self.dough_type = dough_type

    def calculate_price(self):
        base_price = get_base_price(self.size)
        dough_price = get_dough_price(self.dough_type)
        return base_price + dough_price

class Order:
    def __init__(self, pizza):
        self.pizza = pizza

    def calculate_price(self):
        pizza_price = self.pizza.calculate_price()
        return pizza_price + pizza_price * 0.1

pizza = Pizza("large", "thick")
order = Order(pizza)
order_price = order.calculate_price()
print(order_price)
```

ChatGPT can detect the code smell within the given code, as the question was clearly asked about this specific code smell. ChatGPT provides one good possible solution by calling the calculate_price method of the Pizza class once and store the result in a pizza_price variable

### 3.1.2.4 Middle-Man

Class or method acts as an intermediary between two other classes but adds little or no value to the overall design.

```
PROMPT:
this code suffers from Middle-Man, can you fix it?

code class Pizza:
    def __init__(self, size, dough_type):
        self.size = size
        self.dough_type = dough_type
    def calculate_price(self):
        base_price = get_base_price(self.size)
        dough_price = get_dough_price(self.dough_type)
        return base_price + dough_price

class Order:
    def __init__(self, pizza):
        self.pizza = pizza
    def get_price(self):
        return self.pizza.calculate_price()

class PizzaShop:
    def __init__(self, order):
        self.order = order
    def get_order_price(self):
        return self.order.get_price()

pizza = Pizza("large", "thick")
order = Order(pizza)
pizzashop = PizzaShop(order)
order_price = pizzashop.get_order_price()
print(order_price)

ChatGPT:
Yes, there is a middle man issue in this code. A middle man occurs when an object simply forwards a message to another object without
adding any value or functionality.

In this case, the PizzaShop class is acting as a middle man between the Order class and the code that needs to calculate the order price.
The PizzaShop class simply forwards the get_price message to the Order class without adding any additional functionality or value.

To fix this issue, you could eliminate the PizzaShop class and simply call the get_price method of the Order class directly. Here's an
example::

class Pizza:
    def __init__(self, size, dough_type):
        self.size = size
        self.dough_type = dough_type

    def calculate_price(self):
        base_price = get_base_price(self.size)
        dough_price = get_dough_price(self.dough_type)
        return base_price + dough_price

class Order:
    def __init__(self, pizza):
        self.pizza = pizza

    def get_price(self):
        return self.pizza.calculate_price()

pizza = Pizza("large", "thick")
order = Order(pizza)
order_price = order.get_price()
print(order_price)
```

ChatGPT can detect the code smell within the given code, as the question was clearly asked about this specific code smell. ChatGPT provides one good possible solution by eliminating the PizzaShop class and simply call the get_price method.

### 3.1.3 Change preventers

These smells refer to code structures that can prevent or impede future changes to the software. Change preventers make it difficult to modify or extend the software, leading to decreased maintainability, increased technical debt, and higher costs. Examples of change preventers include hardcoding values, using global variables, relying on specific implementations rather than abstractions, and implementing complex conditional statements that are difficult to modify. These code smells are important to identify and address during software development to ensure the software can adapt to changing requirements and remain maintainable over time.

#### 3.1.3.1 Shotgun Surgery

A single change is made to multiple classes simultaneously.

```
PROMPT:
this code suffers from Shotgun Surgery, can you fix it?

is there any shotgun surgery in this code class Pizza:
    def __init__(self, size, dough_type):
        self.size = size
        self.dough_type = dough_type
    def calculate_price(self):
        base_price = get_base_price(self.size)
        dough_price = get_dough_price(self.dough_type)
        return base_price + dough_price

class Order:
    def __init__(self, pizza):
        self.pizza = pizza
    def change_pizza(self, size=None, dough_type=None):
        if size:
            self.pizza.size = size
        if dough_type:
            self.pizza.dough_type = dough_type
        self.pizza.calculate_price()

pizza = Pizza("large", "thick")
order = Order(pizza)
order.change_pizza(size="medium")
order_price = order.pizza.calculate_price()
print(order_price)

ChatGPT:
there is a Shotgun Surgery issue in this code. Shotgun Surgery is a code smell where making a change requires modifying multiple classes
because the responsibility for a particular behavior is spread across multiple classes.

In this case, if we want to make a change to the way pizza sizes are handled (e.g., if we want to add additional size options), we would
need to modify both the Pizza and Order classes. Specifically, we would need to add new size options to the Pizza class, and then update
the change_pizza method of the Order class to handle the new size options.

To fix this issue, we could move the responsibility for handling pizza sizes entirely into the Pizza class, and eliminate the change_pizza
method from the Order class. Here's an example:

class Pizza:
    def __init__(self, size, dough_type):
        self.size = size
        self.dough_type = dough_type
        self.base_price = get_base_price(self.size)

    def calculate_price(self):
        dough_price = get_dough_price(self.dough_type)
        return self.base_price + dough_price

    def set_size(self, size):
        self.size = size
        self.base_price = get_base_price(self.size)

pizza = Pizza("large", "thick")
order = Order(pizza)
pizza.set_size("medium")
order_price = order.pizza.calculate_price()
print(order_price)
```

ChatGPT can detect the code smell within the given code, as the question was clearly asked about this specific code smell. ChatGPT provides one good possible solution by moving the responsibility for handling pizza sizes entirely into the Pizza class.

### 3.1.3.2  Divergent Change

Many changes are made to a single class.

```
PROMPT:
This code suffers from Divergent Change, can you fix it?

class Pizza:
    def __init__(self, size, dough_type):
        self.size = size
        self.dough_type = dough_type

    def calculate_price(self):
        base_price = get_base_price(self.size)
        dough_price = get_dough_price(self.dough_type)
        return base_price + dough_price

def set_pizza_size(pizza, size):
    pizza.size = size

def set_pizza_dough(pizza, dough_type):
    pizza.dough_type = dough_type

pizza = Pizza("large", "thick")
set_pizza_size(pizza, "medium")
set_pizza_dough(pizza, "thin")

ChatGPT:

Yes, this code suffers from the divergent change code smell because if there is a change in how the size or the dough type affects the
price, then both the Pizza class and the set_pizza_size/set_pizza_dough functions will need to be modified, which can lead to a lot of
duplicated code and maintenance issues. The size and dough_type attributes of the Pizza class are being modified by two external functions,
set_pizza_size and set_pizza_dough. This creates code smells, as these two functions are tightly coupled to the Pizza class, which means
that any changes to the Pizza class will likely require changes to these functions

To avoid divergent change by encapsulating the logic related to changing the pizza size and dough type within the Pizza class itself. By
encapsulating the logic within the Pizza class, we avoid having to modify external functions every time we make a change to the class.
This helps to reduce the risk of errors and ensures that the class is easier to maintain in the long run.

class Pizza:
    def __init__(self, size, dough_type):
        self.size = size
        self.dough_type = dough_type

    def calculate_price(self):
        base_price = get_base_price(self.size)
        dough_price = get_dough_price(self.dough_type)
        return base_price + dough_price

    def set_size(self, size):
        self.size = size

    def set_dough_type(self, dough_type):
        self.dough_type = dough_type

pizza = Pizza("large", "thick")
pizza.set_size("medium")
pizza.set_dough_type("thin")
```

ChatGPT can detect the code smell within the given code, as the question was clearly asked about this specific code smell. ChatGPT provides one good possible solution by encapsulating the logic related to changing the pizza size and dough type within the Pizza class itself.

### 3.1.3.3  Parallel Inheritance Hierarchies

Parallel Inheritance Hierarchies is a code smell in object-oriented programming where two or more classes or interfaces share common behavior or properties but are implemented in separate inheritance hierarchies.

---

**PROMPT:**
This code suffers from **Parallel Inheritance Hierarchies**, can you fix it?

```
class Pizza:
    def __init__(self, size):
        self.size = size

class ToppingDecorator(Pizza):
    def __init__(self, pizza, topping):
        super().__init__(pizza.size)
        self.pizza = pizza
        self.topping = topping

    def get_price(self):
        return self.pizza.get_price() + get_topping_price(self.topping)

class CrustDecorator(Pizza):
    def __init__(self, pizza, crust_type):
        super().__init__(pizza.size)
        self.pizza = pizza
        self.crust_type = crust_type

    def get_price(self):
        return self.pizza.get_price() + get_crust_price(self.crust_type)

pizza = Pizza("large")
pizza = ToppingDecorator(pizza, "mushrooms")
pizza = CrustDecorator(pizza, "stuffed")
print(pizza.get_price())
```

**ChatGPT:**
Yes, the code is using inheritance for the wrong purpose. **The ToppingDecorator and CrustDecorator classes should not inherit from the Pizza class because they do not share the same behavior as the Pizza class**. They should instead be implemented as separate classes that have a reference to a Pizza object and add the topping or change the crust type using **composition rather than inheritance**.

This can be done by **creating a Pizza class with a get_price() method, and then creating separate Topping and Crust classes that each have a reference to a Pizza object and a get_price() method that adds the cost of the topping or crust to the price of the pizza**. This would eliminate the need for inheritance and make the code more flexible and easier to maintain.

The code can be fixed by using composition instead of inheritance. Here's an example:

```
class Pizza:
    def __init__(self, size):
        self.size = size
        self.toppings = []
        self.crust_type = "regular"

    def add_topping(self, topping):
        self.toppings.append(topping)

    def set_crust_type(self, crust_type):
        self.crust_type = crust_type

    def get_price(self):
        base_price = get_base_price(self.size)
        toppings_price = sum(get_topping_price(topping) for topping in self.toppings)
        crust_price = get_crust_price(self.crust_type)
        return base_price + toppings_price + crust_price

pizza = Pizza("large")
pizza.add_topping("mushrooms")
pizza.set_crust_type("stuffed")
print(pizza.get_price())
```

In this example, we have a single Pizza class that contains methods for adding toppings and changing the crust type. Instead of creating separate classes for each type of decoration, we simply call the appropriate methods on the Pizza object. This allows for greater flexibility and simplicity in the code.

---

ChatGPT can detect the code smell within the given code, as the question was clearly asked about this specific code smell. The ToppingDecorator and CrustDecorator classes should not inherit from the Pizza class because they do not share the same behavior as the Pizza class. Thus, ChatGPT provides one good possible solution by creating a Pizza class with a get_price() method, and then creating separate Topping and Crust classes that each have a reference to a Pizza object and a get_price() method that adds the cost of the topping or crust to the price of the pizza.

### 3.1.4 Dispensables

These smells refer to the parts of the code that are not necessary and can be removed without affecting the overall functionality of the program. These code smells are usually harmless but can have an impact on the maintainability of the code.

#### 3.1.4.1 Lazy class

A place holder class that initializes its resources or properties until they are needed.

```
PROMPT:
This code suffers from Lazy class, can you fix it?

class Pizza:
    def __init__(self, size, toppings, price):
        self.size = size
        self.toppings = toppings
        self.price = price

    def __str__(self):
        return f"size:{self.size} toppings:{self.toppings} price:{self.price}"

pizza = Pizza("large", ["mushrooms"], 15.99)
print(pizza)

ChatGPT:
Yes, in this code, it's possible that this class could be considered a lazy class if it doesn't have any meaningful behavior or if its
behavior could be combined with another class. Refactoring a lazy class would involve removing the class if it is not needed. However,
if the class is needed, it can be refactored by adding methods or functionality to it that would make it more useful and eliminate the
"lazy" aspect. Here is an example of how the Pizza class can be refactored to add more functionality:

class Pizza:
    def __init__(self, size, toppings):
        self.size = size
        self.toppings = toppings

    def add_topping(self, topping):
        self.toppings.append(topping)

    def remove_topping(self, topping):
        self.toppings.remove(topping)

    def get_price(self):
        base_price = get_base_price(self.size)
        toppings_price = sum(get_topping_price(topping) for topping in self.toppings)
        return base_price + toppings_price

    def calculate_discounted_price(self, discount):
        base_price = get_base_price(self.size)
        toppings_price = sum(get_topping_price(topping) for topping in self.toppings)
        return (base_price + toppings_price) * (1 - discount)

pizza = Pizza("large", ["mushrooms"])
pizza.add_topping("olives")
print(pizza.get_price())
print(pizza.calculate_discounted_price(0.2))

In this refactored version of the Pizza class, two methods have been added: add_topping and remove_topping, which allow toppings to be
added or removed from the pizza. Additionally, a calculate_discounted_price method has been added to calculate the price of the pizza
after applying a discount. These changes eliminate the "lazy" aspect of the original class, making it more useful and functional
```

ChatGPT can detect the code smell within the given code, as the question was clearly asked about this specific code smell. The pizza class doesn't have any meaningful behavior. Thus, ChatGPT provides one good possible solution by creating four methods. Another solution was to eliminate the class if it is not need and introduce an object.

*3.1.4.2 Data class*

A data class refers to a class that contains only fields for accessing them and does not have any behavior.

```
PROMPT:
This code suffers from Data Class, can you fix it?

class Pizza:
    def __init__(self, size, toppings, price):
        self.size = size
        self.toppings = toppings
        self.price = price

    def __str__(self):
        return f"size:{self.size} toppings:{self.toppings} price:{self.price}"

pizza = Pizza("large", ["mushrooms"], 15.99)
print(pizza)

ChatGPT:
Yes, in this code, the Pizza class only has attributes that represent the size, toppings, and price of a pizza, and a method that returns
a string representation of the pizza.

While there is nothing inherently wrong with creating a data class, it may be more appropriate to create a simple data structure, like a
named tuple or a dictionary, to represent the pizza. Alternatively, if the class needs to have behavior or methods, then it should be
designed accordingly.

To fix this, we removed the price parameter from the constructor, and added a calculate_price method that calculates the price of the
pizza based on its size and toppings. We also added an apply_discount method that takes a discount as input and calculates the discounted
price of the pizza.

The __str__ method still exists, but now it calls calculate_price to display the price, ensuring that the displayed price is always up-
to-date.

By adding behavior to the class, we have eliminated the data class code smell.

class Pizza:
    def __init__(self, size, toppings):
        self.size = size
        self.toppings = toppings

    def calculate_price(self):
        base_price = get_base_price(self.size)
        toppings_price = sum(get_topping_price(topping) for topping in self.toppings)
        return base_price + toppings_price

    def apply_discount(self, discount):
        self.price = self.calculate_price() * (1 - discount)

    def __str__(self):
        return f"size:{self.size} toppings:{self.toppings} price:{self.calculate_price()}"
```

ChatGPT can detect the code smell within the given code, as the question was clearly asked about this specific code smell. The pizza class doesn't have any meaningful behavior. ChatGPT provides one possible removed the price parameter from the constructor, and added a calculate_price method that calculates the price of the pizza.

### 3.1.4.3 *Duplicate code*

Code that are repeated multiple times within a program or across different programs.

```
PROMPT:
This code suffers from Duplicate code, can you fix it?

class Pizza:

    def __init__(self, size):
        self.size = size
        self.toppings = []

    def add_topping(self, topping):
        self.toppings.append(topping)

    def remove_topping(self, topping):
        self.toppings.remove(topping)

    def get_price(self):
        base_price = get_base_price(self.size)
        toppings_price = sum(get_topping_price(topping) for topping in self.toppings)
        return base_price + toppings_price

    def calculate_discounted_price(self, discount):
        base_price = get_base_price(self.size)
        toppings_price = sum(get_topping_price(topping) for topping in self.toppings)
        return (base_price + toppings_price) * (1 - discount)

pizza = Pizza("large")
pizza.add_topping("mushrooms")
print(pizza.get_price())

ChatGPT:
Yes, in this code, the code suffers from duplicate code. The get_price() and get_discounted_price() methods both calculate the base price
and toppings price in the same way. This can lead to maintenance issues if there are changes that need to be made to the calculation of
the price.

To fix this, the common code for calculating the price can be extracted into a separate method that both get_price() and
get_discounted_price() can call. This avoids the duplication of code and makes the code more maintainable.

Here's an example of how the code can be refactored:

class Pizza:
    def __init__(self, size):
        self.size = size
        self.toppings = []

    def add_topping(self, topping):
        self.toppings.append(topping)

    def remove_topping(self, topping):
        self.toppings.remove(topping)

    def calculate_price(self):
        base_price = get_base_price(self.size)
        toppings_price = sum(get_topping_price(topping) for topping in self.toppings)
        return base_price + toppings_price

    def get_price(self):
        return self.calculate_price()

    def get_discounted_price(self):
        return self.calculate_price() * 0.8

pizza = Pizza("large")
pizza.add_topping("mushrooms")
print(pizza.get_price())
print(pizza.get_discounted_price())
```

ChatGPT can detect the code smell within the given code, as the question was clearly asked about this specific code smell. The pizza class doesn't have any meaningful behavior. ChatGPT can detect the duplicate code as the get_price() and get_discounted_price() methods both calculate the base price and toppings price in the same way. Thus, ChatGPT extracts the common code for calculating the price into a separate method that both get_price() and get_discounted_price() can call.

*3.1.4.4   Dead code*

Code that is never executed during the runtime of a program

```
PROMPT:
This code suffers from Dead code, can you fix it?

class Pizza:

    def __init__(self, size):
        self.size = size
        self.toppings = []

    def add_topping(self, topping):
        self.toppings.append(topping)

    def remove_topping(self, topping):
        self.toppings.remove(topping)

    def get_price(self):
        base_price = get_base_price(self.size)
        toppings_price = sum(get_topping_price(topping) for topping in self.toppings)
        return base_price + toppings_price

    def calculate_discounted_price(self, discount):
        base_price = get_base_price(self.size)
        toppings_price = sum(get_topping_price(topping) for topping in self.toppings)
        return (base_price + toppings_price) * (1 - discount)

pizza = Pizza("large")
pizza.add_topping("mushrooms")
print(pizza.get_price())
```

**ChatGPT:**
Yes, in this code, If **the method calculate_discounted_price is intentionally not being used and is intended to be kept for future use**, then there is no need to fix it. However, if it is indeed dead code, it can be **removed** to improve code clarity and maintainability.

Here's an updated version of the code with the calculate_discounted_price method removed:.

```
class Pizza:
    def __init__(self, size):
        self.size = size
        self.toppings = []

    def add_topping(self, topping):
        self.toppings.append(topping)

    def remove_topping(self, topping):
        self.toppings.remove(topping)

    def get_price(self):
        base_price = get_base_price(self.size)
        toppings_price = sum(get_topping_price(topping) for topping in self.toppings)
        return base_price + toppings_price

pizza = Pizza("large")
pizza.add_topping("mushrooms")
print(pizza.get_price())
```

ChatGPT can detect the code smell within the given code, as the question was clearly asked about this specific code smell. The pizza class doesn't have any meaningful behavior. Thus, ChatGPT provides one good possible solution by removing the uncessary method.

.

### 3.1.4.5 Speculative generality

Coding a feature that it will be used for a future intended use in a way that is not yet known

```
PROMPT:
This code suffers from Speculative Generality, can you fix it?

class Pizza:

    def __init__(self, size, dough_type, toppings):
        self.size = size
        self.dough_type = dough_type
        self.toppings = toppings

     def calculate_price(self):
        base_price = get_base_price(self.size)
        dough_price = get_dough_price(self.dough_type)
        toppings_price = get_toppings_price(self.toppings)
        return base_price + dough_price + toppings_price

    def prepare_for_future_delivery(self, delivery_type):
        raise NotImplementedError("Delivery type not implemented yet")
```
**ChatGPT:**
Yes, in this code, **the prepare_for_future_delivery method in the Pizza class is an example of speculative generality code smell**. This method is not currently being used and is not needed in the current implementation of the class. It is a speculative feature that has been added to the class in case it is needed in the future. This can add unnecessary complexity and clutter to the code. It's better to only add features that are currently needed and refactor the code later if new features become necessary.

In the folowing example, The prepare_for_future_delivery method **is removed** as it doesn't provide any useful functionality at the moment and only serves as a speculative generality. If a need for this method arises in the future, it can be added then.

```
class Pizza:
    def __init__(self, size, dough_type, toppings):
        self.size = size
        self.dough_type = dough_type
        self.toppings = toppings

    def calculate_price(self):
        base_price = get_base_price(self.size)
        dough_price = get_dough_price(self.dough_type)
        toppings_price = get_toppings_price(self.toppings)
        return base_price + dough_price + toppings_price
```

ChatGPT can detect the code smell within the given code, as the question was clearly asked about this specific code smell. The pizza class doesn't have any meaningful behavior. Thus, ChatGPT provides one good possible solution by removing the uncessary method.

### 3.1.4.6  Unnecessary comment

A method is filled with explanatory comments

```
PROMPT:
This code suffers from Unnecessary Comment, can you fix it?

class Pizza:
    def __init__(self, size):
        self.size = size
        self.toppings = []

    def add_topping(self, topping):
        """Add topping to the pizza."""
        self.toppings.append(topping)

    def remove_topping(self, topping):
        """Remove topping from the pizza."""
        self.toppings.remove(topping)

    def get_price(self):
        """Calculate the price of the pizza based on its size and toppings."""
        base_price = get_base_price(self.size)
        toppings_price = sum(get_topping_price(topping) for topping in self.toppings)
        return base_price + toppings_price

pizza = Pizza("large")
pizza.add_topping("mushrooms")
print(pizza.get_price())
```
**ChatGPT:**
Yes, in this code, **the comments provided for the add_topping and remove_topping methods are redundant, as they simply repeat** what the code already does. The method names "add_topping" and "remove_topping" are clear and self-explanatory, making the comments redundant

```
class Pizza:
    def __init__(self, size):
        self.size = size
        self.toppings = []

    def add_topping(self, topping):
        self.toppings.append(topping)

    def remove_topping(self, topping):
        self.toppings.remove(topping)

    def get_price(self):
        """Calculate the price of the pizza based on its size and toppings."""
        base_price = get_base_price(self.size)
        toppings_price = sum(get_topping_price(topping) for topping in self.toppings)
        return base_price + toppings_price

pizza = Pizza("large")
pizza.add_topping("mushrooms")
print(pizza.get_price())
```

ChatGPT can detect the code smell within the given code, as the question was clearly asked about this specific code smell. The pizza class doesn't have any meaningful behavior. Thus, ChatGPT provides one good possible solution by deleting the uncessary comments.

## 3.1.5 Object Oriented Abusers

These smells refer to a tendency to overuse object-oriented features, such as inheritance, polymorphism, and encapsulation, without fully understanding their implications or benefits. Object-Oriented Abusers can make code harder to understand, maintain, and evolve. They can also lead to performance issues and create unnecessary dependencies between components. To address this smell, developers can use refactoring techniques such as simplifying class hierarchies, using composition instead of inheritance, removing unused abstractions, and simplifying complex code.

### 3.1.5.1 Alternative classes with different interfaces

Providing multiple classes that serve the same purpose but have different ways of interacting with them

```
PROMPT:
This code suffers from alternative classes with different interfaces, can you fix it?

class Pizza:
    def __init__(self, size):
        self.size = size

    def get_price(self):
        return get_price(self.size)

class PizzaDelivery(Pizza):
    def deliver(self):
        delivery_fees = calculate_delivery_fees(self.size)
        send_to_delivery(delivery_fees)

class PizzaTakeaway(Pizza):
    def pick_up(self):
        pick_up_location = get_pickup_location(self.size)
        go_to_pickup_location(pick_up_location)

pizza_delivery = PizzaDelivery("large")
pizza_delivery.get_price()
pizza_delivery.deliver()

pizza_takeaway = PizzaTakeaway("large")
pizza_takeaway.get_price()
pizza_takeaway.pick_up()

ChatGPT:

Yes, in this code, The PizzaDelivery and PizzaTakeaway classes both inherit from the Pizza class, but have different interfaces. This can
lead to confusion and make the code harder to understand and maintain. It would be better to define separate classes with clear and
consistent interfaces for each type of pizza delivery method.One approach to fix it is by providing a common interface order() method for
both PizzaDelivery and PizzaTakeaway classes. This allows for polymorphic behavior, where both classes can be treated as a Pizza object
and have the same methods, while also allowing for each class to implement their specific functionality for the order() method.

class Pizza:
    def __init__(self, size):
        self.size = size

    def get_price(self):
        return get_price(self.size)

    def order(self):
        pass

class PizzaDelivery(Pizza):
    def order(self):
        delivery_fees = calculate_delivery_fees(self.size)
        send_to_delivery(delivery_fees)

class PizzaTakeaway(Pizza):
    def order(self):
        pick_up_location = get_pickup_location(self.size)
        go_to_pickup_location(pick_up_location)

pizza_delivery = PizzaDelivery("large")
pizza_delivery.get_price()
pizza_delivery.order()

pizza_takeaway = PizzaTakeaway("large")
pizza_takeaway.get_price()
pizza_takeaway.order()
```

ChatGPT can detect the code smell within the given code, as the question was clearly asked about this specific code smell. The PizzaDelivery and PizzaTakeaway classes both inherit from the Pizza class, but have different interfaces. Thus, ChatGPT provides one good possible solution by building a common interface for both class.

*3.1.5.2  Switch statement*

when a switch statement becomes too large and complex, with many branches and cases, making it difficult to understand and maintain.

```
PROMPT:
This code suffers from Switch Statement, can you fix it?

class Pizza:
    def __init__(self, size):
        self.size = size

    def calculate_price(self):
        if self.size == "small":
            return 5
        elif self.size == "medium":
            return 7
        elif self.size == "large":
            return 10
        else:
            return "Invalid size"

pizza = Pizza("large")
print(pizza.calculate_price())

ChatGPT:

Yes, in this code, the calculate_price method uses a series of if and elif statements to determine the price of the pizza based on its size. This creates a hard-coded dependency between the size and the price, and it can make it difficult to add new sizes or change the pricing in the future.

Here's one way to fix it using a dictionary to map pizza sizes to their prices. This way, instead of using a series of if/elif statements, the code uses a dictionary to look up the price based on the pizza size.

class Pizza:
    def __init__(self, size):
        self.size = size
        self.prices = {"small": 5, "medium": 7, "large": 10}

    def calculate_price(self):
        if self.size in self.prices:
            return self.prices[self.size]
        else:
            return "Invalid size"

pizza = Pizza("large")
print(pizza.calculate_price())
```

ChatGPT can detect the code smell within the given code, as the question was clearly asked about this specific code smell. The calculate_price method uses a series of if and elif statements to determine the price of the pizza based on its size. Thus, ChatGPT provides one good possible solution by using a dictionary to map pizza sizes to their prices.

### 3.1.5.3 Temporary field

Temporary fields get their values (and thus are needed by objects) only under certain circumstances. Outside of these circumstances, they're empty

```
PROMPT:
This code suffers from Temporary field, can you fix it?

class Pizza:
    def __init__(self, size):
        self.size = size
        self.temp = None

    def prepare(self):
        self.temp = get_oven_temperature(self.size)
        prepare_dough(self.size)
        add_toppings(self.size, self.temp)
        self.temp = None

pizza = Pizza("large")
pizza.prepare()
```

**ChatGPT:**
Yes, this code suffers from the Temporary Field code smell. **The temp variable is only used temporarily during the prepare method and serves no other purpose outside of it**. It is set to None after its usage, indicating that it is not needed beyond the method.

A better approach would be to **eliminate the temporary variable by passing the required data to the functions** that need it directly. This would simplify the code and remove the need for a temporary field.

Here's an example of how the code could be refactored to eliminate the temporary field:

```
class Pizza:
    def __init__(self, size):
        self.size = size

    def prepare(self):
        oven_temp = get_oven_temperature(self.size)
        prepare_dough(self.size)
        add_toppings(self.size, oven_temp)

pizza = Pizza("large")
pizza.prepare()
```

ChatGPT can detect the code smell within the given code, as the question was clearly asked about this specific code smell. The temp variable is only used temporarily during the prepare method and serves no other purpose outside of it. Thus, ChatGPT provides one good possible solution by eliminating the temporary variable by passing the required data to the functions.

### 3.1.5.4 Refused bequest

Refused Bequest occurs when a subclass does not need all the methods or fields inherited from its superclass, but does not override or remove them

```
PROMPT:
This code suffers from Refused bequest, can you fix it?

class Pizza:
    def __init__(self, size, dough_type):
        self.size = size
        self.dough_type = dough_type

    def get_price(self):
        return get_price(self.size)

class TunaPizza:
    def __init__(self, size, dough_type, toppings):
        self.size = size
        self.dough_type = dough_type
        self.toppings = toppings

    def get_price(self):
        base_price = get_price(self.size)
        return base_price + len(self.toppings) * 2

    def get_toppings(self):
        return self.toppings

tuna_pizza = TunaPizza("large", "thick", ["tuna", "olives", "onions"])
print(tuna_pizza.get_price())
print(tuna_pizza.get_toppings())

ChatGPT:
Yes, in this code, the TunaPizza class does not inherit from the Pizza class, but instead replicates some of its functionality. This leads
to code duplication and the need to maintain multiple versions of similar code, which is not optimal. It also violates the principle of
inheritance and the DRY (Don't Repeat Yourself) principle.

A better approach would be to create a Pizza class with a get_price method and then create a separate Toppings class that adds the toppings
to the pizza and calculates the additional cost. This separates the concerns of calculating the base price and calculating the price of
toppings, and also makes the code more extensible if additional toppings or types of pizzas need to be added in the future. Here's an
example:
class Pizza:
    def __init__(self, size, dough_type):
        self.size = size
        self.dough_type = dough_type

    def get_price(self):
        return get_price(self.size)

class Toppings:
    def __init__(self, pizza, toppings):
        self.pizza = pizza
        self.toppings = toppings

    def get_price(self):
        return self.pizza.get_price() + len(self.toppings) * 2

    def get_toppings(self):
        return self.toppings

pizza = Pizza("large", "thick")
toppings = Toppings(pizza, ["mushrooms", "peppers"])
print(toppings.get_price())
print(toppings.get_toppings())
```

ChatGPT can detect the code smell within the given code, as the question was clearly asked about this specific code smell. The TunaPizza class does not inherit from the Pizza class, but instead replicates some of its functionality. Thus, ChatGPT provides one good possible solution by creating a Pizza class with a get_price method and then create a separate Toppings class that adds the toppings to the pizza and calculates the additional cost.

## 3.2 Code generation

Code generation, in the context of this paper, refers to the automatic generation of source code from a given natural language instruction. To this end, we provide the ChatGPT prompt with an English instruction for a code we want it to generate and incrementally modify the instruction. The purpose of this experiment is to generate code in an incremental way. Starting from a simple example function and then asking to change it, making it more complex by introducing new features. We want to experience a guided code generation with a gradual increase in complexity.

**First Prompt:**

Here, in the main prompt we ask for the creation of a JavaScript function. We ask for new features in the subsequent interactions.

```
PROMPT:
Write a javascript program that runs in a browser and allow to load a script file dynamically

ChatGPT:
Sure, here's an example of how you can load a script file dynamically in JavaScript:

function loadScript(url, callback) {
  var script = document.createElement("script");
  script.type = "text/javascript";
  script.src = url;
  script.onload = callback;
  document.head.appendChild(script);
}

// Usage example
loadScript("https://example.com/myscript.js", function() {
  console.log("Script loaded successfully");
});

In this code, the loadScript function takes two arguments: the URL of the script file to load, and a callback function that will be
executed once the script has finished loading.

Inside the function, we create a new <script> element and set its src attribute to the URL of the script file. We also set the onload
property of the element to the callback function. Finally, we append the <script> element to the <head> of the document using the
appendChild method.

When the script file has finished loading, the onload event will be triggered and the callback function will be executed. In the example
above, we simply log a message to the console, but you could do anything you like inside the callback function.
```

**Summary of Execution:**

ChatGPT starts with the generation of a simple function, including an example usage and a final explanation of the produced code. The first answer is the following:

Then we ask the ChatGPT, in a step-by-step process to add new features to the loadScript function. We first ask to add a parameter to avoid browser cache loading of the loaded scripts, then we ask to add the possibility to communicate loading errors, then to add a helper function to load a list of scripts, then to allow the URL list to have elements with a relative path and to add checking if the scripts are well specified (as URL). After that we ask for alternative implementations of that last request.

The following shows one of the final versions of the generated code:

The full experiment can be found in the attached document "GenerateLoadScriptFunction.docx"

**Discussion:**

In this experiment there are some elements we consider relevant to be mentioned:

```
function loadScript(url, callback, preventCache) {
  var script = document.createElement("script");
  script.type = "text/javascript";
  if (isValidUrl(url)) {
    script.src = url;
  } else {
    console.error("Invalid script URL:", url);
    return;
  }
  if (preventCache) {
    url += (/\?/.test(url) ? "&" : "?") + new Date().getTime();
  }
  function onScriptLoad() {
    script.onload = script.onerror = null;
    callback(null, script);
  }
  function onScriptError() {
    script.onload = script.onerror = null;
    callback(new Error("Failed to load script: " + url), null);
  }
  script.onload = onScriptLoad;
  script.onerror = onScriptError;
  document.head.appendChild(script);
                                                        }
function isValidUrl(url) {
  var urlPattern = /^(?:(?:https?|ftp):\/\/)(?:\S+(?::\S*)?@)?(?:\d{1,3}\.\d{1,3}\.\d{1,3}\.\d{1,3}|(?:[a-z\xE1-\xFE0-9]+\.)+[a-z\xE1-\xFE]{2,})(?::\d+)?(?:\/[^\s]*)?$/i;
  return urlPattern.test(url);
}
function loadScripts(urls, callback, preventCache) {
  var loaded = 0;
  var errored = false;
  function onScriptLoad() {
    loaded++;
    if (loaded === urls.length) {
      callback(errored);
    }
  }
  function onScriptError() {
    errored = true;
    loaded++;
    if (loaded === urls.length) {
      callback(errored);
    }
  }
```

- We have done the code generation incrementally. This revealed a better process than having to specify all the functionalities at once. Moreover such a gradual process gave us more confidence on the resulting code;
- After each code generation phase, ChatGPT gives a useful explanation on the resulting code, helpful for understanding the changes;
- ChatGPT is able to remember the changes done along the interaction, and correctly update previously defined functions with new requested features;
- Code may contain bugs due to imprecise specifications (prompt) or because of the code generation process. In the example above, when asking for "allow the url list to have elements with a relative path" the exact prompt has been: "*In the loadScripts could you also allow to specify the urls parameter with elements that contains only just the file name (without protocol and path), in that case you use the path of the previous element of the array. Of course the first element of the array should be a complete url*". This prompt led to an implementation where the line setting *prevUrl* is taken from *urls[index-1]*, which fails if two consecutive urls are relative. In general, test of produced code is still necessary to make sure its bug free;

## 3.3 Step-by-Step execution of source code

Developers often use step-by-step execution of source code to understand semantics of code and to also debug. In this section, we investigate the capability of ChatGPT in the executing a source code in step-by-step fashion.

**Setup:**

The purpose of this experiment is to test capabilities of ChatGPT in debugging. We prompt ChatGPT with the source code of a function and ask ChatGPT to perform a step by step execution of it. We want to see how far ChatGPT can interpret each line, following the algorithm, the control-flow-statements (conditinal, loops) and the status of each variables along the execution. The intention here is to have ChatGPT performing tasks that a debugger would do, however, while debugger typically gives the status of a program in a give time, we want to have a full visible history of the execution of the program.

**First Prompt:**

The main prompt gives the source code of a function, asking ChatGPT to do a step by step execution of it, starting from a statement with an indirect call of the given function:

```
PROMPT:
function digitSum(num1, num2) {
  // Convert numbers to strings to access individual digits
  const strNum1 = String(num1);
  const strNum2 = String(num2);

  // Reverse the strings to start with the ones digit
  const reversedStrNum1 = strNum1.split("").reverse().join("");
  const reversedStrNum2 = strNum2.split("").reverse().join("");

  // Pad the shorter number with zeros to match the length of the longer number
  const maxLength = Math.max(strNum1.length, strNum2.length);
  const paddedStrNum1 = reversedStrNum1.padEnd(maxLength, "0");
  const paddedStrNum2 = reversedStrNum2.padEnd(maxLength, "0");

  // Iterate over the digits and add them up
  const sumDigits = [];
  let carry = 0;
  for (let i = 0; i < maxLength; i++) {
    const digitSum = parseInt(paddedStrNum1[i]) + parseInt(paddedStrNum2[i]) + carry;
    carry = Math.floor(digitSum / 10);
    sumDigits.push(String(digitSum % 10));
  }

  // If there's a carry after the last digit, add it to the result
  if (carry) {
    sumDigits.push(String(carry));
  }

  // Reverse the list of digits and join them to get the final result
  return parseInt(sumDigits.reverse().join(""));
}
using the javascript function above run step by step the execution for the following example:

  console.log(digitSum(438, 30));

Display for each step the values of each variables
```

**Summary of Execution:**

ChatGPT started by recalling the statement to be executed and then, step by step, showed the function's execution, line by line, with each variables state:

ChatGPT continued following a correct step by step the execution flow until step 12 and stopping there. After prompting to continue, ChatGPT just continued until the end, going correctly over the different control flow (if, for), showing intermediate values of complex statements and giving the final function result. The full experiment can be found in the attached document "Digit Sum Algorithm Steps.docx"

**Discussion:**

In this example we have noticed several aspects:

- ChatGPT is capable of interpreting a function call, following its flow, taking in consideration the roles of variables, control statements and also the computation of each sub-statement of more complex statements;
- ChatGPT shows knoledge of the semantic of primitive language statements as well as library functions (e.g. functions for array manipulation, string manipulation, …);
- ChatGPT could provide the functionality of a debugger, showing the correct flow and variable values (we verified each variable with an actual debugger);
- Depending on the length of the step by step execution, it is necessary to prompt ChatGPT to continue in order to get to the end of the algorithm;

```
ChatGPT:
Sure! Here are the step-by-step executions for console.log(digitSum(438, 30)), with the values of each variable displayed for each step:

1. const strNum1 = String(num1);
   - num1 is 438
   - strNum1 is "438"
2. const strNum2 = String(num2);
   - num2 is 30
   - strNum2 is "30"
3. const reversedStrNum1 = strNum1.split("").reverse().join("");
   - strNum1 is "438"
   - strNum1.split("") is ["4", "3", "8"]
   - ["4", "3", "8"].reverse() is ["8", "3", "4"]
   - ["8", "3", "4"].join("") is "834"
   - reversedStrNum1 is "834"
4. const reversedStrNum2 = strNum2.split("").reverse().join("");
   - strNum2 is "30"
   - strNum2.split("") is ["3", "0"]
   - ["3", "0"].reverse() is ["0", "3"]
   - ["0", "3"].join("") is "03"
   - reversedStrNum2 is "03"
5. const maxLength = Math.max(strNum1.length, strNum2.length);
   - strNum1 is "438", which has length 3
   - strNum2 is "30", which has length 2
   - maxLength is 3, the maximum of the two lengths
6. const paddedStrNum1 = reversedStrNum1.padEnd(maxLength, "0");
   - reversedStrNum1 is "834"
   - maxLength is 3
   - reversedStrNum1.padEnd(3, "0") is "834"
   - paddedStrNum1 is "834"
...
```

This example shows that ChatGPT could act as a source code debugger, following the algorithm encoded in a source code and showing sub-statements and variables states. In a short investigation we verified that ChatGPT is not actually evaluating language statements with a language interpreter engine (in this case a JavaScript engine), rather it has semantic knowledge of what the statements produces, therefore it is possible to get an incorrect evaluation of a statement.

## 3.4 Bug fixing

Bug fixing of code refers to the removal of known errors in code and the output of a bug-free code. The following describes our experiment with ChatGPT in the context of bug fixing.

In a recent publication related to code analysis and automatic bug fixing [Sobania et Al. 2023] a performance assessment is illustrated by comparing ChatGPT, CoCoNut and standard APR approached on the problem of fixing bugs of different types. The result shows that ChatGPT has comparable performance like ChatGPT or dedicated deep-learning based APR, outperforming standard APR approaches. The finding shows that bug fixing with ChatGPT may still require a dialog between the developer and ChatGPT, in order to guide the fixing process. Here an example from the paper:

**Setup:**

The purpose of this experiment is to test the capabilities of ChatGPT in bug fixing source code. In this example the researcher requests ChatGPT to check a simple program, driving attention of ChatGPT to noticing a possible fault.

**First Prompt:**

The main prompt gives the source code of a function, asking ChatGPT if it contains a bug:

```
PROMPT:
Does this program have a bug? How to fix it?

  def gcd( a, b ):
    if b == 0:
      return a
    else:
      return gcd( a % b, b )
```

**Summary of Execution:**

ChatGPT at first did not find any bug in the function. After the researcher explained that the program was generating an error, then ChatGPT could spot the but, give an explanation for it and propose a possible solution:

```
ChatGPT:

This is the Euclidean algorithm for finding the greatest common divisor (GCD) of two
integers, a and b. It appears to be implemented correctly. If you are experiencing
issues with this program, it is likely that the problem lies else where in your code.
Without more information about the bug you are experiencing, it is difficult to say
how to fix it.

PROMPT:
No, the function is not correct. E.g., for the input a = 4 and b = 5 I get the following
error message: "Recursion Error: maximum recursion depth exceeded in comparison".

ChatGPT:
The issue is that the function calls itself recusrively, and in some cases, it never
Reaches the base case (b == 0), causing it to recurse indefinitely. The base case of
The gcd algorithm is when the remainder is zero, but in this simple implementation, it's when
B is zero. For example, when a = 4 and b = 5, gcd( 4, 5 ) is called, which then calls
gcd( 4 % 5, 5 ) which is gcd( 4 , 5 ). This causes an infinite loop.

You can fix this by swapping the parameters in the recursive call.

  def gcd( a , b ):
    if a == 0:
      return b
    else:
      return gcd( b % a, a )

This way, if a is zero, the function returns b (which is the gcd of a and b), otherwise
it calls itself with the new values of b % a and a.
```

The paper shows few other examples and gives a description of a comparison study where ChatGPT has been compared with other methods on the base of the QuixBug dataset. The experiments and evaluations can be found in [Sobania et Al. 2023].

**Discussion:**

This example shows that ChatGPT can be used in the scope of finding and fixing bugs on source programs. It may be necessary to guide ChatGPT in the search process, however, it shows abilities of localising errors and suggest solutions for them. At the current state it may be questionable if the performance in bugfixing are in general sufficient to justify the necessary interaction with the users.

## 3.5 Improving code comprehension

Code comprehension is an important software engineering problem that aims at improving the understandability of source code there by improving maitainablity. The goal of this experiment is to check the capability of ChatGPT in the context of Code Comprehension.

**Setup:**

Developers routinely go through hundreds of lines of code to understand the meaning of code. A manual effort in this regard is to focus only on the relevant parts of the code and ignore the remaining. For our experiment, we provide a piece of Python code and ask ChatGPT to refactor it with goal of improving the code comprehension.

**Discussion:**

```
PROMPT:
Improve Code comprehension the following code:
  def foo(n, d):
    v = v + 1
    c = c + 1
    return n + d
```

```
ChatGPT:
The given code has some issues with variable scope and naming, so here is the refactored code:
def foo(num1, num2):
    # Increment variables num1 and num2 by 1
    num1 += 1
    num2 += 1
    # Return the sum of num1 and num2
    return num1 + num2

In this refactored code, I have used more descriptive variable names (assuming that n and d represent num1 and num2 respectively). I have also fixed
the issue with variable scope, since v and c were not defined in the original code. Finally, I have added comments to explain what the function is
doing, and I have indented the lines of code inside the function to make it more readable.
```

As mentioned earlier, a manual effort for code comprehension ignores the irrelevant parts of code. For the given code snippet, a developer is likely to ignore the lines 'v = v + 1' and 'c = c + 1' since these variables are not used in the return value of the function and does not contribute to the overall meaning of computation by the function. When produced with this code to ChatGPT, it refactors the code but changes the semantics of the code. From the comprehension perspective, it adds some natural language comments but since the refactoring is not semantics preserving, we conclude that such refactoring are still beyond the scope of ChatGPT.

## 3.6 Bad coding practices detection

Bad coding practices refer to coding bad practices that do not necessarily cause errors but might lead to problems in the future.

**Setup:**

We provide ChatGPT with a JavaScript code with a commonly known bad code practice and ask it to find it. The particular bad code practice we have in mind is the use of '==' operator for comparing string types in JavaScript.

```
PROMPT:
Find the code smell in the following JavaScript code:
str1 = "Hello"
str2 = "Mello"

if (str1 == str2) {
console.log("Hi");
}
```

**Discussion:**

Being a very common code bad practice, ChatGPT unsurprisingly does not have any problem in finding it. In fact, it also provides a correct solution for it.

```
ChatGPT:
The code smell in the given JavaScript code is the use of the equality operator (==) for string comparison. The equality operator performs type coercion,
which means that if the types of the operands are different, JavaScript will try to convert one of the operands to the type of the other operand. This
can lead to unexpected results and bugs in the code.

For string comparison, it is better to use the strict equality operator (===), which compares the values as well as the types of the operands. So, the
code can be rewritten as:

if (str1 === str2) {
  console.log("Hi");
}
```

## 3.7  Code analysis and summarization

Code summarization refers to the process of generating a natural language explanation of source code. Such summary particularly aids in code understandability and maintainability. ChatGPT has some capability in this regard as evident from the following experiment.

**Setup:**

```
PROMPT:
Summarize the following code:
  def add_nums(num1, num2):
    v = num1 + num2
    return  v
```

We ask the ChatGPT system to summarize a given Python code snippet. The system responds by generating an English language summary of the provided code.

```
ChatGPT:
This code defines a function called "add_nums" that takes in two parameters called "num1" and "num2". The function computes the sum of the two input
numbers and assigns it to a variable "v". Then, the function returns the value of "v". In other words, this function calculates the sum of two numbers
and returns the result.
```

**Discussion:**
To our query about code summarization, the response provided by ChatGPT adequately explains the code. Based on the names of variables, 'num1' and 'num2', ChatGPT system expects them to be numbers at runtime and explains that the function calculates the sum of two numbers.

```
PROMPT:
Summarize the following code:
  def add(val1, val2):
    v = val1 + val2
    return  v
```

As a follow-up exercise, we provide ChatGPT with another code summarization task with minor changes in the code. Instead of 'add_num', the function is called 'add' and we also change the name of the variables from 'num1', 'num2' to 'val1' and 'val2' respectively. The primary template of the produced summary is same for both code examples. With minor changes in the names of the identifiers, ChatGPT system rightly points about that the "the input values can be of any type that supports the "+" operator".

```
ChatGPT:
This code defines a function called "add" that takes in two parameters called "val1" and "val2". The function computes the sum of the two input values
and assigns it to a variable "v". Then, the function returns the value of "v". In other words, this function calculates the sum of two values and
returns the result. The input values can be of any type that supports the "+" operator.
```

# 4     Limitations

In this section we analyze the current limitations of LLM systems in order to give a more comprehensive view on the boundaries of such AI infrastructures [Borji 2023].

## *4.1     Coding Limitations*

### 4.1.1   Limited support for complex logic

As an AI language model, ChatGPT is designed to generate text based on patterns and relationships it has learned from large amounts of data. While ChatGPT can generate simple lines of code based on specific templates or examples it has seen, it may have limitations when it comes to creating complex logic in software development. From knowledge perspective, ChatGPT can only consider the information it has seen in its training data and may not have the context or background knowledge to understand the nuances of complex logic. Moreover, it does not have a deep understanding of software engineering principles, such as design patterns or object-oriented programming, which are often used to develop complex software systems. It cannot also make test or debug the code it generates with a programming interpreter or engine, which means it may produce code with errors or bugs that are difficult to identify or fix. Overall, while ChatGPT may be able to generate simple code snippets, it may not be suitable for generating complex software logic that requires a deep understanding of software engineering principles and creative problem-solving.

### 4.1.2   Limited knowledge of programming languages

ChatGPT may not have knowledge of all programming languages or may have limited knowledge of certain languages, which can limit its ability to generate code for those languages. From a training perspective, its training data do not include a wide range of programming languages and their syntax. This means that ChatGPT's understanding of programming languages may be limited to the patterns and relationships it has learned from the programming language data it was trained on. Moreover, programming languages are constantly evolving, with new versions and updates being released frequently. It can be challenging to keep up with these changes, and it may not be feasible to retrain ChatGPT with updated language syntax and features.

### 4.1.3   Difficulty with Math and Science Questions

ChatGPT has difficulty with math and science questions for several reasons. It has not been specifically designed or trained to understand and answer math and science questions. Math, logic and science can be highly complex subjects that involve a deep understanding of abstract concepts, equations, and formulas. ChatGPT struggles to understand and apply these concepts accurately in its responses [Azaria 2022] [Borji 2023]. Such topics often involve visual information, such as diagrams and equations, which can be difficult for a text-based model like ChatGPT to interpret and understand. Finally as mentioned before, answering mathematical questions require a deep understanding of the physical world and common-sense knowledge, which ChatGPT may not possess to a sufficient extent.

### 4.1.4   Difficulty to manipulate quantitative information (measures, probabilities, …)

There are several reasons why LLMs have difficulty in representing quantitative information. They operate by representing words and phrases as discrete tokens in a vocabulary. Representing continuous numbers as discrete tokens can result in a loss of precision and accuracy. Manipulating and representing quantitative information often involves mathematical operations, that cannot be performed accurately in a discrete language model. Representing quantitative information often requires the ability to perform unit conversions, such as converting between different units of length, mass, or time. This can also be difficult in a discrete language model, as it may not have the ability to understand the relationships between different units of measurement.

## 4.2  *Performance Limitations*

### 4.2.1  Limited Short-Term Memory

ChatGPT architecture lacks explicit memory. Unlike human brains, which have an explicit memory system that allows us to store and recall information over time, ChatGPT does not have such a memory system. The working memory that LLMs can use is limited to the number of past tokens that his sliding window can handle which is in the order of 2000 tokens in ChatGPT. This means that it cannot store information from one conversation turn to the next, and must rely solely on the information that is provided to it in each turn. Moreover, fine-tuning ChatGPT on specific conversational data can help to improve its ability to generate responses in the context of a specific task or domain. However, fine-tuning can also introduce limitations, such as a lack of generalization to new conversational flows and a reduced ability to handle more long-term memories. The data used to train ChatGPT is largely unstructured and often lacks clear conversational flow and context which also contributes to its difficulty to generate coherent and consistent responses in the context of a conversation.

### 4.2.2  Vulnerability to Adversarial Input

ChatGPT can be easily manipulated or misled by malicious or misleading inputs. This is a common issue in many machine learning systems and is particularly relevant in the context of natural language processing and chatbots, where the input text can be easily crafted to trick the system into generating incorrect or inappropriate responses [Borji 2023]. Adversarial inputs can take many forms, including intentionally misleading or confusing statements, statements that are factually incorrect or inconsistent with the context, and statements that are designed to prompt the system to generate specific, undesired responses.

### 4.2.3  Dependence on Large Amounts of Computing Power

ChatGPT requires large amounts of computing power to run its inference network, making it challenging to deploy in resource-constrained environments. Currently it is provided as a free cloud service. Recently (February 10, 2023) it is available for professionals as a low latency paid cloud service.

### 4.2.4  Incapacity to Learn New Data On-line

LLMs up to now are trained offline and used on-line for multiple tasks but are incapable of learning new content on the fly. It is important to note that although the scientific litterature use the terms "zero shot learning", or "few shot learning" to describe the number of examples provided in the prompt to perform a task, there is no learning at all involved in such experiments: No parameters (weights) of the model are modified during such tests.

# 5  Broader Impacts Hazard Analysis

## 5.1  *Overreliance*

Using code generation models can be risky, as there is a tendency to over-rely on the generated outputs. This is especially problematic when ChatGPT provides solutions that only appear correct, but do not perform the task as intended. This risk is particularly high for novice programmers, and in certain contexts, could have significant safety implications. Therefore, it's critical to exercise human oversight and vigilance when using code generation systems.

To promote safe use of these systems, it's essential to provide clear documentation that highlights the limitations of the models. Additionally, empirical investigations should be conducted to determine how to ensure vigilance in practice, considering different levels of user experience, UI designs, and tasks. One challenge that researchers need to consider is the possibility of "automation bias" as the capabilities of these systems improve. This could make it increasingly difficult to guard against the tendency to blindly rely on the generated outputs without verifying their accuracy.

## 5.2 Misalignment

As an AI language model, ChatGPT is trained to generate code that is like its training distribution, using a next token prediction objective. However, this can result in unhelpful suggestions for the user, even though the model has the capability to be more helpful. For example, if the user has subtle mistakes in their code, ChatGPT may suggest code that appears good but is ultimately incorrect.

This is known as an alignment failure, where the model is not aligned with the user's intentions. When a system is able to perform a task, but chooses not to, it is considered misaligned. In contrast, a system that fails to perform a task due to incompetence is not misaligned. Misalignment is an important problem to study because it is likely to become worse as the capabilities of AI systems increase. For instance, the trend of scaling model size indicates that misalignment would persist and even worsen if data, parameters, and training time were increased.

Although the misaligned behavior of current models may not cause significant harm, it could become more dangerous and harder to eliminate as model capabilities increase. A highly capable but misaligned model trained on user approval might produce obfuscated code that appears good to the user but is undesirable or harmful. Therefore, it is important to address misalignment in AI systems to ensure that they are safe, effective, and aligned with human values and intentions. This requires ongoing research and development of alignment techniques, as well as regular evaluation of model behavior to detect and correct misaligned behavior.

## 5.3 Bias

Like other language models trained on Internet data, ChatGPT is susceptible to manipulation, which can result in racist, denigratory, and otherwise harmful outputs, such as code comments. This highlights the need for interventions to address these issues. Code generation models, in general, pose additional bias and representation concerns beyond natural language. ChatGPT can generate code with structures that reflect stereotypes about gender, race, emotion, class, the structure of names, and other characteristics.

This issue is especially problematic in the context of users who may rely too heavily on ChatGPT or use it without thoroughly thinking through project design, as it could have significant safety implications. To address these concerns, it is important to discourage over-reliance on the model and implement mitigation strategies such as filtration or modulation of generated outputs, as well as providing appropriate documentation to users.

Overall, it is essential to address the bias and representation issues that arise from the use of code generation models like ChatGPT, to create more inclusive and equitable technologies. This requires ongoing research and development of techniques to identify and mitigate these issues, as well as education and awareness-raising efforts to promote responsible use of these technologies.

## 5.4 Cybersecurity

ChatGPT's ability to generate vulnerable or misaligned code could have various impacts on the security landscape. It is crucial for qualified operators to review and validate the generated code before executing or trusting it, especially without appropriate precautions. While future code generation models could potentially produce more secure code than the average developer, it is uncertain at this time.

There is a risk that ChatGPT could be misused to aid cybercrime. ChatGPT as more powerful code generation models are developed, further research into mitigations and continued study of model capabilities will be necessary. The non-deterministic nature of ChatGPT and similar systems could enable more advanced malware, particularly through techniques like generating polymorphic malware. While application security and model deployment strategies, such as rate-limiting access and abuse monitoring, can manage this threat in the near term, the effectiveness of these mitigations may not scale linearly as more capable models are developed.

Additionally, ChatGPT models can learn patterns present in their training data, including sensitive data present in source code. As ChatGPT is trained on public repositories, any sensitive data present in the training data should already be considered compromised. Therefore, the public data should generally be treated as untrusted, as attackers could potentially corrupt training data to trigger specific model behaviors at runtime.

Overall, it is crucial to consider the potential security implications of code generation models like ChatGPT and take appropriate measures to mitigate any risks, including human review, filtration, monitoring and usage of threat modeling tools [Shi et al. 2021].

## 5.5  *Environment*

ChatGPT, like other large generative models, has a significant energy footprint from both training and inference. The training and the fine-tunning process consumes hundreds of petaflops/s days of compute. These high energy demands have important environmental implications, and it is crucial to consider them when deploying these models. It is worth noting that the training was performed on a platform (Azure) that purchases carbon credits and sources a significant amount of renewable energy, which helps to reduce its carbon footprint. However, the compute consumption of large models like ChatGPT also has costs in the wider supply chain that can be concentrated in certain regions.

In the long term, as code generation becomes more prevalent and challenging problems require significant inference, the compute demands of these models could grow even larger than ChatGPT's training. Therefore, it is essential to consider the energy implications of these models and explore ways to reduce their environmental impact, such as developing more efficient algorithms, using renewable energy sources, and incorporating energy efficiency metrics into the model design process.

## 5.6  *Economy*

The impact of code generation and associated capabilities on the economy and labor market is complex and multifaceted. While ChatGPT, at its current capability level, may lead to some cost savings by increasing programmer productivity, it is important to note that engineers engage in various tasks other than coding, such as collaborating with colleagues, writing design specifications, and upgrading existing software stacks. Therefore, the overall impact on productivity may be limited.

Furthermore, ChatGPT's tendency to import packages at different rates could result in an uneven distribution of benefits, favoring some package authors over others, particularly if programmers and engineers rely heavily on ChatGPT's suggestions. In the long term, the effects of code generation technologies on software-related labor markets and the broader economy could be more significant as capabilities improve. Therefore, further research is needed to fully understand the economic and labor market impacts of these technologies and to develop appropriate responses to mitigate any negative consequences.

## 5.7  *Legislation*

Generated code raises various legal considerations, including the fair use of training data. While training AI systems on public GitHub repositories has been previously deemed as fair use, the study found that ChatGPT models rarely generate code that is identical to the contents of training data. When the generated code appears identical to the training data, it is due to the model's predictive weightings rather than the retention and copying of specific code.

In addition, generated code is responsive and customizable to the user's input, and the user maintains complete control over editing and accepting the generated code. This makes code generation similar to auto-suggest or auto-completion features found in other authorship tools such as document editors, where the finished work is still seen as the authors. Nonetheless, further legal considerations may arise as code generation technology continues to evolve, and it will be important to continue monitoring these developments.

# 6  Summary, Conclusions and Recommendations

This paper has screened the capabilities and limitations of ChatGPT, a large language model designed for use in software development. ChatGPT's power lies in its ability to act as an intermediary between the programmer's natural language and the machine code. As a result, it offers a solution that simplifies the development process for non professional programmers. This streamlines the development process for non-professional programmers and simplifies the task of coding.

OpenAI released upgraded GPT-3 and Codex models in March 2022. These models had impressive new capabilities and were trained on a large dataset. In November 2022, OpenAI introduced the "GPT-3.5" series, including the text and code models, with further fine-tuning. OpenAI also released ChatGPT based on GPT-3.5 models. The tested Codex models's ability to solve programming problems, was found to outperformed GPT models with up to 28.8% accuracy. Fine-tuning on correctly implemented functions resulted in Codex-S, which could solve 37.7% of problems with a single sample. Codex-S generated at least one correct function within 100 samples for 77.5% of the problems.

However, it is important to consider that the current version of ChatGPT is better tuned, therefore its accuracy is higher than the shown figures within the studied paper.

Furthermore, this paper provides a detailed examination of ChatGPT's potential applications in software system engineering. The analysis highlights several areas where ChatGPT can be effectively applied, including code smells detection, refactoring, generation, step-by-step execution of source code, bug fixing, code comprehension improvement, and code analysis summarization. These applications offer significant benefits to software developers, helping them to write more efficient, maintainable, and comprehensible code. The findings of this paper suggest that ChatGPT has the potential to transform the field of software engineering and streamline the development process for non-professional programmers.

ChatGPT also has several limitations in coding, performance, and text generation. In terms of coding, ChatGPT struggles with complex logic due to a lack of context and background knowledge. The constantly evolving nature of programming languages limits ChatGPT long term coding capablities. It also has difficulty answering math and science questions, manipulating quantitative information, and representing continuous numbers as discrete tokens. In terms of performance, ChatGPT has limited short-term memory and is vulnerable to adversarial input, making it challenging to deploy in resource-constrained environments. ChatGPT is also incapable of learning new content on the fly, and although it is available as a free cloud service and as a paid low latency cloud service for professionals, it requires large amounts of computing power to run its inference network. Text generation limitations include bias in training data and instability of responses when doubts are raised, which may lead to a reformulation of the response with opposite affirmations, or overconfidence in sticking to its responses despite user corrections.

The Broader Impacts Hazard Analysis of ChatGPT highlights several concerns with using code generation models. The overreliance on the generated outputs of ChatGPT can be risky, especially when the model provides solutions that only appear correct but do not perform the intended task. Misalignment is another concern, where ChatGPT may suggest unhelpful code to the user due to its training distribution. The susceptibility to bias and representation issues is also a significant problem, and mitigation strategies such as filtration and providing appropriate documentation are essential. The security implications of ChatGPT generating vulnerable or misaligned code need to be considered, and appropriate measures must be taken to mitigate risks. The high energy demands of ChatGPT during training and inference have important environmental implications, and it is crucial to explore ways to reduce their environmental impact. Additionally, the impact of ChatGPT on the economy and labor market is complex and multifaceted, and further research is needed to fully understand the economic and labor market impacts of these technologies. Lastly, legal considerations such as fair use of training data need to be taken into account as code generation technology continues to evolve.

ChatGPT has the potential to be a valuable tool in software development in any organization. ChatGPT limitations are understood and its output is supervised and improved. It can be used in a variety of applications such as writing code for PhD projects, refactoring old projects code, and translating programming languages after implementation, acting as a coding buddy etc. In particular,by incorporating ChatGPT as a tool for buddy programming, non-professional programmers could receive automated suggestions for code implementation, refactoring, and language translation, improving their code quality and development skills. This could ultimately lead to reduced overallsoftware development costs, improved software quality, and scalability in software development. Early tests of model driven development (MDD) shows that ChatGPT could also be used in this direction and based on the experience of our institute in this domain [Ceravola et Al. 2006][Sadik et Al. 2019], MDD with ChatGPT will certainly be investigated more systematically in the near future.

The limitations and impact of ChatGPT are critical research areas. One such area is security, as ChatGPT does not currently consider security issues when generating code. However, we can anticipate that future models of ChatGPT will be trained on code that emphasizes security, which will enable them to write more secure code than the average programmer. Another important research direction is designing language models that can be trained with minimal energy consumption, as training and fine-tuning large language models require a significant amount of energy. In addition, ChatGPT's inability to evolve in tandem with evolving programming languages can be addressed through continuous learning or learning on the fly. By doing so, ChatGPT's scalability can be improved, and it can keep up with the fast pace of programming language development.

In conclusion, ChatGPT is a powerful tool that can be utilized for software architecturing, whether in software design using forward engineering methods or in refactoring existing software using reverse engineering techniques. In this sense it is a perfect tool for collaborative intelligence [Sendhoff & Wersing 2020] and can/should be used with human developers in the loop. By incorporating ChatGPT into research and innovation strategy, innovative approaches can be explored and advancements in software development achieved, leading to an increase in both work scale and work quality. The potential of ChatGPT in this field is vast and exciting new developments can be expected as we continue to integrate it into software practices.

# 7 Acknowledgment

# 8 References

Azaria A. (2022) ChatGPT Usage and Limitations. hal-03913837.

Bird, C., Ford, D., Zimmermann, T., Forsgren, N., Kalliamvakou, E., Lowdermilk, T., & Gazit, I. (2022). Taking Flight with Copilot: Early insights and opportunities of AI-powered pair-programming tools. Queue, 20(6), 35-57.

Borji, A. (2023). A Categorical Archive of ChatGPT Failures. arXiv preprint arXiv:2302.03494.

Ceravola, A., Joublin, F., Dunn, M., Eggert, J., Stein, M., & Goerick, C. (2006, October). Integrated research and development environment for real-time distributed embodied intelligent systems. In 2006 IEEE/RSJ International Conference on Intelligent Robots and Systems (pp. 1631-1637). IEEE.

Chollet, F. (2017). The limitations of deep learning. Deep learning with Python.

Chen, M., Tworek, J., Jun, H., Yuan, Q., Pinto, H. P. D. O., Kaplan, J., ... & Zaremba, W. (2021). Evaluating large language models trained on code. arXiv preprint arXiv:2107.03374.

Gozalo-Brizuela, R., & Garrido-Merchan, E. C. (2023). ChatGPT is not all you need. A State of the Art Review of large Generative AI models. arXiv preprint arXiv:2301.04655.

Joublin, F., Ceravola, A., Deigmoeller, J., Gienger, M., Franzius, M., & Eggert, J. (2023). A Glimpse in ChatGPT Capabilities and its impact for AI research. arXiv preprint arXiv:2305.06087.

OpenAI, T. B. (2022). Chatgpt: Optimizing language models for dialogue. OpenAI. https://openai.com/blog/chatgpt/

Sadik, A. R., Goerick, C., & Muehlig, M. (2019, December). Modeling and simulation of a multi-robot system architecture. In 2019 International Conference on Mechatronics, Robotics and Systems Engineering (MoRSE) (pp. 8-14). IEEE.

Schuurmans, D. (2023). Memory Augmented Large Language Models are Computationally Universal. arXiv preprint arXiv:2301.04589.

Shi, Z., Graffi, K., Starobinski, D., & Matyunin, N. (2021). Threat modeling tools: A taxonomy. IEEE Security & Privacy, 20(4), 29-39.

Sendhoff, B., & Wersing, H. (2020, September). Cooperative intelligence-a humane perspective. In 2020 IEEE International Conference on Human-Machine Systems (ICHMS) (pp. 1-6). IEEE.

Sobania, D., Briesch, M., Hanna, C., & Petke, J. (2023). An Analysis of the Automatic Bug Fixing Performance of ChatGPT. arXiv preprint arXiv:2301.08653.

van Dis, E. A., Bollen, J., Zuidema, W., van Rooij, R., & Bockting, C. L. (2023). ChatGPT: five priorities for research. Nature, 614(7947), 224-226.

Zaremba, W., & Brockman, G. (2021). OpenAI ChatGPT. OpenAI. https://openai.com/blog/openai-ChatGPT/